\documentclass[12pt]{article}
\usepackage{epsf}
\usepackage{amsmath}
\usepackage{graphics}
\usepackage{cite}

\renewcommand{\thefootnote}{\#\arabic{footnote}}
\begin{document}

\newcommand{\gtrsim}{ \mathop{}_{\textstyle \sim}^{\textstyle >} }
\newcommand{\lesssim}{ \mathop{}_{\textstyle \sim}^{\textstyle <} }

\newcommand{\rem}[1]{{\bf #1}}

\renewcommand{\thefootnote}{\fnsymbol{footnote}}
\setcounter{footnote}{0}
\begin{titlepage}

\def\thefootnote{\fnsymbol{footnote}}

\begin{center}
%\hfill hep-th/yymmnnn\\

\vskip .5in
\bigskip
\bigskip
{\Large \bf Upper and Lower Bounds on Gravitational Entropy }

\vskip .45in

{\bf Paul H. Frampton$^{(a)}$\footnote{frampton@physics.unc.edu}
and Thomas W. Kephart$^{(b)}$\footnote{tom.kephart@gmail.com}}

\bigskip
\bigskip

{$^{(a)}$ Department of Physics and Astronomy, UNC-Chapel Hill, NC 27599.}

{$^{(b)}$ Department of Physics and Astronomy, Vanderbilt University,
Nashville, TN 37235.}

\end{center}

\vskip .4in

\begin{abstract}
The gravitational entropy of the Universe is large
and subtle to calculate. A
lower bound is well known from supermassive
black holes at the centers of galaxies, but the remainder 
is harder to pin down. A parametric model of clumped matter 
entropy is provided. We suggest a new upper bound due to 
dark matter halos, that is far below 
the holographic bound, yet above
that for the supermassive black holes. 
\end{abstract}

\end{titlepage}

\renewcommand{\thepage}{\arabic{page}}
\setcounter{page}{1}
\renewcommand{\thefootnote}{\#\arabic{footnote}}

\newpage

\noindent {\it Introduction.}

\bigskip

As interest grows in pursuing alternatives to the Big Bang,
including cyclic cosmologies, it becomes more pertinent to
address the difficult question of what is the present entropy
of the universe?

\bigskip

Entropy is particularly relevant to cyclicity because it
does not naturally cycle but has the propensity 
to increase monotonically because of the second law of thermodynamics. 
In one recent proposal  
the entropy is jettisoned at turnaround. In any case, for cyclicity to be
possible
there must be a gigantic reduction in entropy, 
at deflation in \cite{BF},
of the visible universe at some time during each cycle.

\bigskip

Standard treatises on cosmology \cite{Tolman, Weinberg}
address the question of the  entropy of the
universe and arrive at a generic formula for a thermalized
gas of the form

\begin{equation}
S = \frac{2 \pi^2}{45} g_{*} V_U T^3 
\label{thermal}
\end{equation}
where $g_{*}$ is the number of degrees of freedom, $T$ is
the Kelvin temperature and $V_U$ is the volume of the visible universe.
From Eq.(\ref{thermal}) with $T_{\gamma}=2.7 $ K and $T_{\nu} = T_{\gamma}
(4/11)^{1/3} = 1.9 $ K we find the
entropy in CMB photons and neutrinos are roughly equal today
\begin{equation}
S_{\gamma}(t_0) \sim S_{\nu}(t_0) \sim10^{88}.
\label{rad}
\end{equation}
Our topic here is the gravitational entropy, $S_{grav} (t_0)$. 
Following the same path as in
Eqs.(\ref{thermal},\ref{rad}) we obtain for a thermal
gas of gravitons $T_{grav} = 0.9 $ K and then
\begin{equation}
S_{grav}^{(thermal)}(t_0) \sim 10^{86}
\label{thermalgrav}
\end{equation}
which is a couple of orders of magnitude below that for
photons and neutrinos. On the other hand, while radiation
thermalizes at $T\sim 0.1$ eV for which the measurement 
of the black body spectrum provides good evidence
and there is every reason, though no direct evidence,
to expect that the relic neutrinos were thermalized at $T \sim 1$ MeV,
the thermal equilibration of the present gravitons is less definite.
If gravitons did thermalize, it was  
at or above the Planck scale, $T \gtrsim 10^{19}$ GeV, when everything is
uncertain because
of quantum gravity effects. If the gravitons are in a non-thermalized
gas their entropy will be lower than in Eq.(\ref{thermalgrav}), for the same
number density. But
there are larger contributions to gravitational entropy from elsewhere.

In this paper we will review upper and lower bounds on the gravitational entropy of the Universe 
and then argue that dark matter halos contain the dominant component. While a detailed model based on halo distributions is not described, we do provide a parametrization of the halo entropy and give a one parameter bound that is sufficient for our purposes. The model is a simplification in that it assumes all halos have equal masses and ignores cluster and supercluster halos. Hence our results will be semi-quantitative. Relaxing these approximations would unduly complicate the model without significantly improving any insight it hopefully provides.

\bigskip

\noindent {\it Upper Limit on Gravitational Entropy}.

\bigskip

We shall assume that dark energy has zero entropy 
 and therefore we must concentrate on the gravitational entropy associated with dark
matter.
\footnote{Our bounds on entropy are consistent with \cite{bekenstein}}
The dark matter is clumped into halos with typical mass
$M(halo) \simeq 10^{11}M_{\odot}$  and radius $R(halo) = 10^4 pc \simeq 3 \times 10^{17} km
\simeq
10^{17} r_S(M_{\odot}$), where $M_{\odot} \simeq 10^{57} GeV \simeq
10^{30}
 kg$
is the solar mass and $r_S(M_{\odot})$ its Schwartzschild radius. There are, say, $10^{12}$
halos in the visible universe
whose total mass is $\simeq 10^{23} M_{\odot}$ and
corresponding Schwarzschild radius is $r_S(10^{23} M_{\odot}) \simeq 3 \times
10^{23} km
\simeq	10$ Gpc. This coincides with the radius of the visible
universe corresponding to the critical density and has led to an upper limit
for the gravitational entropy
of one black hole with mass $M_U = 10^{23} M_{\odot}$.
Using $S_{BH}(\eta M_{\odot}) \simeq 10^{77} \eta^2$ corresponds to the
holographic principle \cite{Hooft,Susskind} for the upper limit
on the gravitational entropy of the visible universe:

\begin{equation}
S_{grav} (t_0) 
\lesssim S_{grav}^{(HOLO)}(t_0) \simeq 10^{123}
\label{HOLO}
\end{equation}.

\noindent which is 37 orders of magnitude greater than
for the thermalized graviton gas in Eq.(\ref{thermalgrav})
and leads us to suspect (correctly) that Eq.(\ref{thermalgrav})
is a gross underestimate. Nevertheless, Eq.(\ref{HOLO})
does provide a credible upper limit, an overestimate yet to be refined downwards
below,
on the quantity of interest, $S_{grav}(t_0)$. 

\bigskip

The reason why a thermalized gas of gravitons grossly underestimates
the gravitational entropy is because of the 'clumping'
effect of gravity on entropy \cite{Penrose}.
Because gravity is universally attractive its entropy
is increased by clumping. This is somewhat counter-intuitive
since the opposite is true for the more familiar ideal gas. It is
best illustrated by the fact that a black hole always has
maximal entropy by virtue of the holographic principle.
Although it is difficult to estimate gravitational entropy
we will attempt to be semi-quantitative in implementing the idea.

\bigskip

Let us consider one halo with mass $M(halo) = 10^{11} M_{\odot}$ and
radius $R_{halo} = 10^{17} r_S(M_{\odot}) \simeq 10^4 pc$. Applying
the holographic principle with regard to the clumping effect would
give an overestimate for the halo entropy $S_{halo} (t_0)=S_{halo}^{(HOLO)} (t_0)
$ which we propose to estimate with a phenomenological clumping factor
by the following ansatz,

\begin{equation}
S_{halo} (t_0) = S_{halo}^{(HOLO)} (t_0)  \left( \frac{r_S(halo)}{R(halo)}
\right)^p
\label{clumping}
\end{equation}
where $p$ is a real parameter. Since $r_S (halo) \leq R(halo)$,
Eq.(\ref{clumping}) ensures that $S_{halo} \leq S_{halo}^{HOLO}(t_0)$
provided that $p \geq 0$.
Actually the holographic principle requires that $S_{halo} \leq
S_{BH}(M_{halo})$
and since $S_{BH} \propto r_S^2$, this requires that $p \geq 2$
in Eq.(\ref{clumping}).

\bigskip

The value $p=2$ provides a more realistic upper limit on the
present gravitational entropy of the universe
$S_{grav}(t_0)$ than Eq.(\ref{HOLO}). Using our average
values for $M_{halo}$ and $R_{halo}$ and a number $10^{12}$
of halos this gives

\begin{equation}
S_{grav}(t_0) < 10^{110}
\label{upperlimit}
\end{equation}
which is many orders of magnitude below the holographic limit
of Eq.(\ref{HOLO}). The physical reason is that the clumping
to one black hole is very incomplete as there are a trillion
disjoint halos. If all the halos coalesced to one
black hole, and there is no reason
to expect this given the present accelerating expansion of the universe,
the entropy would reach the maximum value
in Eq.(\ref{HOLO}) of $10^{123}$ but at present the upper limit
is given by Eq. (\ref{upperlimit}).

\bigskip

\noindent {\it Lower Limit on Gravitational Entropy}.

\bigskip

It is widely believed that most, if not all, galaxies
contain at their core a supermassive black hole with
mass in the range $10^5 M_{\odot}$ to $10^9 M_{\odot}$
with an average mass of about $10^7 M_{\odot}$. In
our simplified model we assume all have this average mass, so each  
carries an entropy 
$S_{BH} ({\rm supermassive}) \simeq 10^{91}.$
Since there are $10^{12}$ halos this provides the lower limit
on the gravitational entropy \cite{Penrose}, \cite{KN} of

\begin{equation}
S_{grav} (t_0) \gtrsim 10^{103}
\label{lowerlimit}
\end{equation}
which together with Eq.(\ref{upperlimit}) provides a
seven
order of magnitude window for $S_{grav} (t_0)$.

The lower limit in Eq.(\ref{lowerlimit}) from the
galactic supermassive black holes may be the largest contributor
to the entropy of the present universe but this seems to us
highly unlikely because they are so very small, occupying 
$\sim 10^{-36}$ of the volume.
 Each supermassive
black hole is about the size of our solar system or smaller
and it seems counterintuitive that essentially
all of the entropy is so concentrated. On the other hand, since 
gravitational entropy grows with clumping, this could be the case.

As already mentioned, gravitational entropy is associated with the clumping
of matter because of the long range unscreened nature
of the gravitational force. This is why we propose that
the majority of the entropy is associated with the 
largest clumps of generalized matter: the dark matter halos associated
with galaxies and clusters of galaxies.

\bigskip

\noindent {\it Most Likely Value of Gravitational Entropy}.

\bigskip

In the phenomenological formula for clumping, Eq.(\ref{clumping}),
the parameter $p$ must satisfy $2 \leq p < \infty$ because for $p=2$
the halo entropy is as high as it can be, being equal to that of
the largest single black hole into which it could collapse, while
for $p \rightarrow \infty$, the halo has no gravitational entropy beyond
that of the supermassive black hole at its core.
 Thus, our upper and lower limits 

\begin{equation}
10^{110} \geq S_{grav} (t_0) \geq 10^{103}
\label{limits}
\end{equation}
correspond to $p=2$ and $p \rightarrow \infty$ in Eq.(\ref{clumping})
respectively.
We may include the supermassive black holes
in Eq.(\ref{clumping}) by noticing \cite{footnote} that
$S_{grav}(t_0) = 10^{(124 - 7p)}$ and therefore, from Eq.(\ref{limits}),
$2 \leq p \leq 3$.

\bigskip

Actually, the power $p$ in Eq.(\ref{clumping}) must depend on the 
halo radius $R_{halo}$ such that $p(R_{halo}) \rightarrow 2$ as
$R_{halo} \rightarrow r_S$, the Schwarzschild radius, when the
halo collapses to a black hole. For the present non-collapsed 
status of the halos, $ p > 2 $ is necessary since the black
hole represents a maximum possible entropy. One might also expect  $p$ to be
dependent on density and therefore radially dependent, but we assume this is mild
enough to allow us to obtain order of magnitude estimates by employing
constant $p$.

 \bigskip

We believe the pursuit of better understanding
of gravitational entropy in clumps of matter
with mass above $M_e = 10^{21}$ kg. (see Eq. (\ref{Me}) below)
may provide
a very fruitful approach towards a satisfactory theory of quantum
gravity. The gravitational entropy
we are discussing, if it exists, may well be a quantum mechanical phenomenon.

\bigskip

We can apply the same considerations based on Eq.(\ref{clumping})
to gravitation within a single star like the Sun. The Sun has $(r_S/R_{\odot})
\sim 10^{-5}$
and with $p_a = 3$ we find $S_{\odot}^{(grav)} \sim 10^{72}$, 
far above the standard	$S_{\odot} \sim 10^{57}$, suggesting
a contribution from stars 
to the gravitational entropy
of about $\sim 10^{95}$.

\bigskip

As the gravitating object we consider becomes smaller 
the relative importance of gravitational entropy to
non-gravitational entropy changes. Let us obtain a rough
estimate of the mass $M_e$ at which the two contribution
are comparable.

\bigskip

Suppose $M_e = \eta M_{\odot} \simeq 10^{30} \eta$ kg. and so we
wish to determine $\eta$. We can estimate $\eta$ by the fact that the
gravitational
entropy in Eq.(\ref{clumping}) is not linear in $\eta$ but has
a different dependence. Let us take a typical density 
of the putative object to be
$\rho = 5\rho_{H_20} = 5 \times 10^{12} kg/(km)^3$. The radius
of a sphere with mass $M_e$ is then $R \simeq 4 \times 10^5 \eta^{1/3}$ km.
Thus the gravitational entropy from Eq.(\ref{clumping}) is

\begin{equation}
S_{grav} = (10^{77} \eta^2) \left( \frac{3 \eta}{4 \times 10^5 \eta^{1/3}}
\right) \simeq 10^{72} \eta^{8/3}
\label{Sgrav}
\end{equation}

\noindent revealing the $\eta$ dependence of $S_{grav}$.
On the other hand, the non-gravitational entropy may be
estimated by counting baryons to give the usual form linear in $\eta$

\begin{equation}
S_{non-grav} \simeq 10^{57} \eta.
\label{Snongrav}
\end{equation}

\bigskip

The two contributions, $S_{grav}$ of Eq.( \ref{Sgrav} ) and $S_{non-grav}$
of Eq. ( \ref{Snongrav} )
become comparable when
$\eta^{- 5/3} \sim 10^{15}$ or $ \eta \sim 10^{-9}.$
This equality mass $M_e$ is
about 
\begin{equation}
M_e \simeq 0.1\% M_{\oplus} \simeq 10^{21} {\rm kg.}
\label{Me}
\end{equation}

\bigskip

\noindent If we consider much smaller
masses such as a baseball ($\sim 1$ kg) or a primordial black
hole with lifetime comparable to the age of the universe
($\sim 10^{12}$ kg), the gravitational entropy becomes totally
negligible.

\bigskip

According to our phenomenological clumping ansatz, Eq.(\ref{clumping}),
the entropy of solar system objects can be larger
than conventionally assumed, the Sun by $10^{15}$, the Earth by
$10^5$. We possess no derivation of this gravitational entropy component
and publish this idea only to prompt
more mathematically rigorous arguments to estimate the 
contribution of gravitational clumping to entropy.

\bigskip

Another reason to suspect a large gravitational entropy outside of black
holes
comes from considering the gravitational collapse of an 
object of mass, say, $M = 10 M_{\odot}$ which contains
$\sim 10^{58}$ nucleons
and hence non-gravitational entropy
$S \sim 10^{58}$.
Under gravitational collapse, it is conventionally believed that
the entropy gradually increases, though {\it not} by orders of magnitude,
as the radius decreases toward the Schwarzschild radius.
When the trapped surface of a black hole   appears,
the entropy jumps to $\sim 10^{79}$, an increase
of some twenty orders of magnitude.
While not excluded, this is intuitively implausible.
On the other hand,
with the clumping factor of Eq.(\ref{clumping}) and the 
starting density we have employed of
$\rho = 5 \rho_{H_2O}$, the starting
entropy from Eq.(\ref{Sgrav}) is already $\sim 10^{72 + 8/3}
\sim 5 \times 10^{74}$, and less dramatic entropy increase is needed. 
In fact, $S_{halo}$ is a smooth function as $R \rightarrow r_S$.

\bigskip

There is a second consideration which provides circumstantial
evidence for unsuspected gravitational entropy. If, as in Eq.(\ref{lowerlimit}),
the cosmological entropy is dominated by the supermassive black
holes, it implies that almost all the entropy is confined
to a trillion objects each of radius $\sim 10^{-6}$ pc occupying
$\sim 10^{-33}$ of the halo volume.
Altogether they compose only $\sim 10^{-36}$ of the total volume
of the visible universe. Although not excluded by any
deep principle, it is disconcerting.

\bigskip

Let us attempt to make a somewhat more quantitative argument 
out of the idea of how entropy grows with gravitational clumping.
At last scattering density perturbations in the dark matter were small,
$\frac{\delta\rho}{\rho} \sim 10^{-5},$
but today there are regions where
$\frac{\delta\rho}{\rho} \sim 1$
where we expect the gravitational entropy has increased enormously even though
the entropy in photons per comoving volume has remained relatively constant. 

\bigskip

The non-clumped component
of the universe expands adiabatically.
How do we get the entropy of a clump? Assume the dark matter is in the
form of very
light particles. For a clump of size $L_{gal} = 10^4$ pc, the lightest
particles that can clump
are of mass $m \sim 10^{-20}$ eV. Otherwise their associated Compton wavelength 
may be too large.
Recall the galactic mass is
$M_{gal} \sim 10^{12}M_{\odot} \sim 10^{69}$GeV.
If this is all in dark matter (ignore baryons, etc.), then there
can be at most
$N \sim \frac{M_{gal}}{m} \sim 10^{98}$
dark matter particles in a halo, or about
$N_{U} \sim 10^{110}$
dark matter particles in the universe that are now clumped.

\bigskip

If the dark matter particles start off at rest (similarly to nonthermal
axions) but then start to fall into
clumps, we can argue that their degrees of freedom get excited, 
i.e., as the particles fall into the potential well they gain kinetic energy. So
these gravitational
d.o.f.s give approximately zero contribution to the total entropy
before density
perturbations start to grow, but they now contribute $\sim 10^{110}$. If the
masses of the dark matter particles are larger, the contribution to the entropy
will be proportionally smaller. The mass $m \sim 10^{-20}$ eV
provides an approximate
upper bound on the gravitational entropy. The lower bound for the entropy
in this particular approach
would be very small if the dark matter particles are far heavier such as
WIMPs at the TeV scale.

\bigskip

\noindent {\it Discussion and Conclusions}

\bigskip

Entropy is always a subtle concept, nowhere more so than for
gravity. This is why we are bold enough to make such
approximate
estimates of the present gravitational entropy of the visible
universe.
Our results are concerned only with orders of magnitude
and we hope our upper and lower limits 
$10^{110}$ and $10^{103}$ are credible \cite{pre}.

\bigskip

These already show that the universe's entropy is dominated
by gravity, being at least 13 orders of magnitude above
the known entropies, each $\simeq 10^{88}$,
for photons and relic neutrinos.

\bigskip

Using the clumping idea 
and an heuristic clumping factor
dependent on a parameter $p$ 
suggests that the gravitational entropy
is dominated not by the well known galactic supermassive black holes 
which contribute $\simeq 10^{103}$ but
by a larger, possibly much larger, contribution
from the dark matter halos which can provide  
not more than (for $p \rightarrow 2$)
about $10^{110}$ which is still
many orders of magnitude below
the holographic bound $\simeq 10^{123}$.
Our estimates for the entropy of the universe
have an error of one order of magnitude.

\bigskip

Everything we have said about gravitational entropy due to clumping may be
nonsensical but, if it exists,
it is reasonable to expect it 
to be non-classical and an effect of quantum gravity like
the holographic bound and the black hole entropy. Since string theory
has had some success in those two cases, it could help 
in deciding whether our speculations are idle. More
optimistically, the study of gravitational entropy
will itself lead to a better theory of quantum gravity, hopefully
the correct one.

\bigskip

If our speculations are correct: the contribution of radiation to
the total entropy of the Universe is less than 1 part in $10^{16}$; supermassive
black holes at galactic cores contribute more but still
less, possibly much less, than  a few per cent of the total; 
we propose that the gravitational entropy
contained only in stars is already greater than the entropy
of electromagnetic radiation;
and the gravitational entropy contained in dark matter halos can be
the biggest contributor to the entropy of the universe \cite{catchall}.

\begin{center}

{\bf Acknowledgments}

\end{center}

\bigskip

This work was supported by U.S. Department of Energy grants number
DE-FG02-06ER41418 (PHF) and  DE-FG05-85ER40226 (TWK).


\begin{thebibliography}{100}
\bibitem{footnote} Taking the log of Eq. (6) and holding everything but $p$ fixed we can write
$$\log_{10} S = a +bp  $$ or equivalently $S=10^{a+bp}$. Now requiring
$a$ and $b$ be chosen to satisfies the two boundary conditions
set by Eq. (10) gives the result. Note that $p \rightarrow \infty$ 
corresponds to no additional gravitational entropy from the dark matter.
\bibitem{pre} These bounds are implicit in the discussions of  \cite{Penrose}
but they were neither explored in the detail we have provided here, nor were they modeled.
The lower bound was also pointed out in \cite{KN}.
\end{thebibliography}
\end{document}